\documentclass[twocolumn,superscriptaddress,aps,pra]{revtex4}
\usepackage{amsfonts}
\usepackage{amsmath}
\usepackage{amssymb}
\usepackage{graphicx}
\usepackage{color}
\usepackage{ulem}
\usepackage{dcolumn}
\usepackage{bm}
\usepackage{url}
\usepackage[colorlinks, urlcolor = cyan, citecolor = blue, linkcolor = magenta]{hyperref}

\begin{document}

\title{Generalized Parametric Resonance in a Spin-1 Bose-Einstein Condensate}
\author{Peng Xu}
% \email{2014202020003@whu.edu.cn}
\affiliation{Institute for Advanced Study, Tsinghua University, Beijing, 100084, China}
\author{Wenxian Zhang}
% \email{wxzhang@whu.edu.cn}
\affiliation{School of Physics and Technology, Wuhan University, Wuhan, Hubei 430072, China}
\date{\today}

\begin{abstract}
We propose a generalized Mathieu's equation (GME) which well describes the dynamics for two different models in spin-1 Bose-Einstein condensates. The stability chart of this GME differs significantly from that of Mathieu's equation and the unstable dynamics under this GME is called generalized parametric resonance. A typical region of $\epsilon \gtrsim 1$ and $\delta \approx 0.25$ can be used to distinguish these two equations. The GME we propose not only explains the experimental results of Chapman's group [Nat.Commun.7,11233(2016)] in nematic space with a small driving strength, but predicts the behavior in the regime of large driving strength. Besides, the model in spin space we propose, whose dynamics also obeys this GME, can be well tuned such that it is easily implemented in experiments.
\end{abstract}

\maketitle

\section{Introduction}
\label{sec:1}

\begin{figure}[t]
    \centering
    \includegraphics[width = 3.5in]{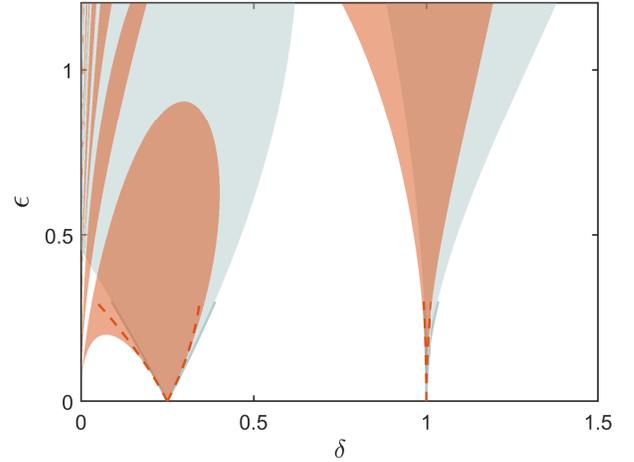}
    \caption{The chart of unstable regions for the ME (light cyan (light gray)) and the GME (orange (dark gray)). The shaded areas show the numerical results and the lines depict the transition between stable and unstable regions based on the perturbation theory.}
    \label{fig:1}
\end{figure}

Parametric resonance dates back to Mathieu's work in 1868, where he considered an elliptic membrane~\cite{Mathieu1868Memoire}. Since then, it has been applied in many fields such as quantum mechanics~\cite{Ruby1996Applications, Alberti2009Engineering, Ma2011Photon, Chen2011Controlling, New2011Introduction, Nation2012Stimulating} and general relativity~\cite{Zlatev1998Parametric, Easther2000Gravity, Fedichev2004Cosmological, Nation2012Stimulating}. Parametric resonance is different from direct excitation. In principle, parametric resonance is periodically tuning some parameters of the system, while direct excitation is simply periodically kicking the oscillator but keeps its intrinsic properties intact. Generally speaking, the way to tune the parameters can be classified into two classes, first in systems with periodic forcing, such as vibration of a string by periodically varying tension~\cite{Thomsen2003Vibrations}; second in stability studies of periodic motions in nonlinear autonomous systems, such as ``particle in the plane'' problem~\cite{Yang1967Vibrations, Yang1968Forced}. In the past two decades, as the cold atom physics develops, parametric resonance plays an important role in this field, where it can be used to measure the trap frequency, to control the superfluid-Mott insulator transition~\cite{Stoferle2004Transition}, and to excite the collective density modes~\cite{Jin1996Collective, Mewes1996Collective, Engels2007Observation, Jaskula2012Acoustic, Clark2017Collective}.

Except for Mathieu's equation (ME) describing parametric resonance, there exist many other generalized equations, such as the ones including damping, delay or quasiperiod, etc~\cite{Kovacic2018Mathieu}. The unstable regions show more complicated and interesting patterns~\cite{Cao2019Mathieu}. In 2016, Chapman's group studied the dynamics of populations on $|f = 1, m_f = 0 \rangle$ in nematic space of a spin-1 Bose-Einstein condensate (BEC) by applying a periodically driving second order Zeeman energy~\cite{Hoang2016Parametric}. The populations on $|0\rangle$ fraction exhibit parametric resonance in the regime of small driving strength. However, it is intrinsically described by the generalized Mathieu's equation (GME),
\begin{align}
    \ddot{x} + \left[ \delta + \epsilon \cos(t) + \frac{\epsilon^2}{4 \delta} \cos^2(t) \right] x = 0,
    \label{eq:1}
\end{align}
which we propose in this paper. The stability chart of Eq.~(\ref{eq:1}) is shown in Fig.~\ref{fig:1}. When the driving parameter $\epsilon$ is very small, the unstable regions of ME (light cyan regions) and GME (orange regions) almost overlap except that the unstable regions of the GME slightly move to left comparing with the unstable regions of the ME. However, when $\epsilon$ is large or $\delta$ is very small, the stability chart is significantly different from that of the ME. For example, when $\epsilon \gtrsim 1$ and $\delta \approx 0.25$, the region is unstable for GME while it is stable for ME. The detailed calculations and discussions for the stability chart of Eq.~(\ref{eq:1}) can be found in the Appendix.~\ref{sec:appa}.

In this paper, by employing this GME we not only explain the experimental results of Chapman's group in nematic space with a small driving strength, but also predict the behavior in the regime of large driving strength. Furthermore, we propose another model in spin space of a spin-1 BEC, which can be described well by the GME; more importantly, the parameters of this model are well tuned such that it is easily implemented in experiments.

The paper is organized as follows. In Sec.~\ref{sec:2}, we describe the system of a spin-1 BEC. In Sec.~\ref{sec:3}, we map two models into the GME. In Sec.~\ref{sec:4}, we benchmark that the dynamics under the Hamiltonian definitely obeys the GME. The conclusions are given in Sec.~\ref{sec:5}. More details about the GME are discussed in the Appendices.

\section{Hamiltonian of a spin-1 BEC}
\label{sec:2}

We consider a trapped dipolar spin-1 BEC whose Hamiltonian is~\cite{Ho1998Spinor, Law1998Quantum, Yi2004Quantum, Yi2006Spontaneous, Yi2006Magnetization},
\begin{align}
    \hat{H} = \hat{H}_0 + \hat{H}_d,
    \label{eq:2}
\end{align}
where $\hat{H}_d$ represents magnetic dipolar interaction between atoms and $\hat{H}_0$ the rest part. In the second quantized form
\begin{widetext}
\begin{align}
    \label{eq:3}
    \hat{H}_0 &= \int d\bm{r} \hat{\Psi}_{m}^{\dag}(\bm{r})
    \left[ \left( - \frac{\hbar^2 \nabla^2}{2 M} + V(\bm{r}) \right) \delta_{m n} \right] \hat{\Psi}_{n}(\bm{r}) \nonumber \\
    &+ \frac{c_0}{2} \int d\bm{r} \hat{\Psi}_{m}^{\dag}(\bm{r}) \hat{\Psi}_{n}^{\dag}(\bm{r}) \hat{\Psi}_{m}(\bm{r}) \hat{\Psi}_{n}(\bm{r}) + \frac{c_2}{2} \int d\bm{r} \hat{\Psi}_{m}^{\dag}(\bm{r}) \hat{\Psi}_{m'}^{\dag}(\bm{r}) \bm{f}_{m n} \cdot\bm{f}_{m' n'} \hat{\Psi}_{n}(\bm{r})\hat{\Psi}_{n'}(\bm{r}), \\
    \hat{H}_d &= \frac{c_d}{2} \int \frac{d\bm{r} d\bm{r'}}{|\bm{r} - \bm{r'}|^3} \left[\hat{\Psi}_{m}^{\dag}(\bm{r}) \hat{\Psi}_{m'}^{\dag}(\bm{r'}) \bm{f}_{m n} \cdot \bm{f}_{m' n'} \hat{\Psi}_{n}(\bm{r}) \hat{\Psi}_{n'}(\bm{r'}) - 3 \hat{\Psi}_{m}^{\dag}(\bm{r}) \hat{\Psi}_{m'}^{\dag}(\bm{r'}) (\bm{f}_{m n} \bm{\cdot e}) (\bm{f}_{m' n'} \bm{\cdot e}) \hat{\Psi}_{n}(\bm{r}) \hat{\Psi}_{n'}(\bm{r'}) \right], \nonumber
\end{align}
\end{widetext}
where $M$ is the mass of the atom, $V(\bm{r})$ the trapping potential. $\hat{\Psi}_m$ is the field annihilation operator for the spin component $m = - 1, 0, + 1$, $\bm{f} = (f_x, f_y, f_z)$ with $f_{x, y, z}$ being spin-1 matrices, and $\bm{e = (r - r') / |r - r'|}$ a unit vector. The collisional interaction parameters are $c_0 = {4 \pi \hbar^2 (a_0 + 2 a_2)} / {3 M}$ and $c_2 = {4 \pi \hbar^2 (a_2 - a_0)} / {3 M}$ with $a_{0(2)}$ being the $s$-wave scattering length of two spin-1 atoms in the combined symmetric channel of total spin 0(2). The dipolar interaction parameter is $c_d = {\mu_0 g_{F}^2 \mu_{B}^2} / {4 \pi}$ with $\mu_0$ being the vacuum magnetic permeability, $g_F$ the Land$\acute{\rm{e}}$ $g$-factor and $\mu_B$ the Bohr magneton. The repeated indices are summed.

Under the single mode approximation $\hat{\Psi}_m(\bm{r}) \simeq  \phi(\bm{r}) \hat{a}_m$, we can significantly simplify Eq.~(\ref{eq:2}) for the following three experimental situations. First, in a spherical trap where the dipolar interaction becomes negligible, the Hamiltonian in an external magnetic field is simplified as (after dropping some constants)~\cite{Law1998Quantum, Barnett2010Antiferromagnetic, Zhang2013Generation, Xu2019Efficient},
\begin{align}
    \hat{H} = c'_2 \frac{\hat{\bm{L}}^2}{N} - q \hat{a}_0^\dagger \hat{a}_0,
    \label{eq:4}
\end{align}
where $\hat{\bm{L}} = \sum_{m n} \hat{a}_m^\dagger \bm{f}_{m n} \hat{a}_n$, $c'_2 = c_2 N \int d\bm{r} |\phi(\bm{r})|^4 / 2$, $N$ the total particle number, and $q$ the quadratic Zeeman energy. Second, by applying a far off-resonant blue-detuned $\pi$-polarized microwave field, which couples the $F = 1$ manifold to the $F = 2$ manifold, the $SU(2)$ symmetry of the spin-exchange term $\propto c_2$ in Eq.~(\ref{eq:3}) can be broken. In fact, the spin-exchange collisions become energetically forbidden, due to that the coupling coefficient for $| F = 1, M_F = 0 \rangle \rightarrow | F = 2, M_F = 0 \rangle$ transition is much larger than the coupling coefficients for $| F = 1, M_F = \pm 1 \rangle \rightarrow | F = 2, M_F = \pm 1 \rangle$ transitions. Under this condition, the Hamiltonian in a spherical trap is remarkably simplified as~\cite{Sorensen2001Many},
\begin{align}
    \hat{H} = c'_2 \frac{\hat{J}_z^2}{N}, \nonumber
\end{align}
where $\hat{\bm{J}} = \sum_{m n} \hat{a}_m^\dagger \bm{\sigma}_{m n} \hat{a}_n$, $\bm{\sigma}$ the Pauli matrices, $|F = 1, M_F = 1 \rangle$ and $|F = 1, M_F = - 1 \rangle$ representing spin-up and spin-down, respectively. Furthermore, we apply two far off-resonant red-detuned $\sigma_+$ and $\sigma_-$ polarized microwave fields to couple $|F = 1, M_F = - 1 \rangle \rightarrow |F = 2, M_F = 0 \rangle$ and $|F = 1, M_F = 1 \rangle \rightarrow |F = 2, M_F = 0 \rangle$, respectively. Then the effective Hamiltonian becomes,
\begin{align}
    \hat{H} = c'_2 \frac{\hat{J}_z^2}{N} - p \hat{J}_x,
    \label{eq:6}
\end{align}
where the effective linear Zeeman splitting $p \propto \Omega^2 / \Delta$ with $\Omega$ the driving strength of $\sigma_{+, -}$ microwave field, $\Delta$ the detuning between $|F = 1, M_F = \pm 1 \rangle$ and $|F = 2, M_F = 0 \rangle$. Third, in a cylindrical trap where the linear term in the dipolar interaction is negligibly small, the Hamiltonian in a small magnetic field becomes (after dropping some constants)~\cite{Yi2006Magnetization, Xu2017Rebuilding},
\begin{align}
    \hat{H} = c'_d \frac{\hat{L}_z^2}{N} - p \hat{L}_x,
    \label{eq:7}
\end{align}
where $c'_d = (3 c_d N / 4) \int d\bm{r} d\bm{r}' |\phi(\bm{r})\phi(\bm{r}')|^2 ({1 - 3 \cos^{2} \theta}) / |\bm{r}-\bm{r'}|^3$, and $p$ the linear Zeeman energy.

The above three models may be realized in a spin-1 BEC. In general, the typical value of $|c'_2|\sim 7$ Hz in a $^{87}$Rb BEC, and $c'_2 \sim 25$ Hz in a $^{23}$Na BEC. The dipolar interaction $c'_d$ is usually smaller than 1 Hz in these BECs. In Eq.~(\ref{eq:7}) the linear Zeeman energy is $p = 0.7$ MHz at a magnetic field $B_x \approx 1$ G for a $^{87}$Rb atom. The effect of nonlinear term in Eq.~(\ref{eq:7}) may play a role if the transversal field $B_x\sim 1\;\mu$G, which is possible in a magnetic shielding room~\cite{Eto2013Spin, Eto2014Control}. In Eq.~(\ref{eq:6}) the effective linear Zeeman splitting can be controlled by both $\Delta$ and $\Omega$; and in Eq.~(\ref{eq:4}) the quadratic Zeeman energy $q$ is proportional to $72 B^2 \; \text{Hz} / \text{G}^2$ in a $^{87}$Rb BEC, and $277 B^2  \; \text{Hz} / \text{G}^2$ in a $^{23}$Na BEC, where $B$ with Gauss unit is the strength of the external magnetic field. Both $p$ in Eq.~(\ref{eq:6}) and $q$ in Eq.~(\ref{eq:4}) are easily tuned in experiments in a wide range from a value smaller than $c'_2$ to a value much larger than $c'_2$. By varying $p$ and $q$, the effects of the nonlinear terms in these equations are easily observed in experiments. As shown below, the dynamics either depending on Eq.~(\ref{eq:4}) in nematic space or Eq.~(\ref{eq:6}) in spin space, under certain approximations, can be mapped to a driven harmonic oscillator~\cite{Luigi2013Spin}. According to this mapping process under these approximations, we find the physics such as dynamics of these two different models is the same, which significantly simplifies and unifies our understandings for these two models.

\section{Quantum dynamics under the mapped GME}
\label{sec:3}

The Shr$\ddot{o}$dinger equation with the Hamiltonian Eq.~(\ref{eq:6}) is
\begin{align}
    (c'_2 \frac{\hat{J}_z^2}{N} - p \hat{J}_x) \sum_{n = - N}^N c_n(t) |n\rangle &=& i \frac{\partial}{\partial t} \sum_{n = - N}^N c_n(t) |n\rangle,
\end{align}
where $|n\rangle$ is chosen as the eigenstate of $\hat{J}_z$. By multiplying a bra $\langle m|$, the above equation becomes
\begin{align}
    &\frac{c'_2}{N} c_m(t) m^2 - \frac{p}{2} \sqrt{(N - m + 1) (N + m )} c_{m - 1}(t) \nonumber \\
    &- \frac{p}{2} \sqrt{(N + m + 1) (N - m )} c_{m + 1}(t) = i \frac{\partial}{\partial t} c_m(t).
    \label{eq:9}
\end{align}
By choosing $\varepsilon \ll x \ll 1$ with $\varepsilon = 1 / N$ and $x = m / N$ and taking the continuum limit, $c_m(t)$ becomes a continuous function $c(x,t)$. After we neglect high order terms $o(x^2)$ and $o(\varepsilon^2)$, Eq.~(\ref{eq:9}) is reduced to
\begin{align}
    - \frac{1}{2} \frac{p}{N} \frac{\partial^2 c}{\partial x^2} + \frac{1}{2} (p + 2 c'_2) N x^2 c - p N c = i \frac{\partial c}{\partial t}.
    \label{eq:10}
\end{align}
The third term in the left-hand side is nothing but a constant potential and can be neglected. Clearly, this is an equation for a harmonic oscillator. The mass $M$ and the frequency $\omega_0$ of the oscillator are found from $1 / M = p / N$ and $M \omega_0^2 = (p + 2 c'_2) N$. Then the effective Hamiltonian for this oscillator is thus
\begin{align}
    \hat{H} = \frac{\hat{P}^2}{2 M} + \frac{1}{2} M \omega_0^2 \hat{x}^2
    \label{eq:H_eff}
\end{align}
where
\begin{align}
    M = N / p, \;
    \omega_0 = \sqrt{p (p + 2 c'_2)}, \;
    \hat{x} |m\rangle = x |m\rangle.
    \label{eq:11}
\end{align}

According to Eq.~(\ref{eq:H_eff}) and the equation of motion, i.e., $i \langle \dot{\hat{O}} \rangle = \langle [\hat{H}, \hat{O}] \rangle$, we obtain a dynamical equation similar to a classical harmonic oscillator,
\begin{align}
    \langle \ddot{\hat{x}} \rangle + p (p + 2 c'_2) \langle \hat{x} \rangle = 0,
    \label{eq:12}
\end{align}
where $\langle O \rangle$ represents the expectation for a coherent state. When one tunes the effective external magnetic field $p$ as $p [1 + \epsilon_0 \cos(\nu t)]$, the above equation becomes,
\begin{align}
    \langle \ddot{\hat{x}} \rangle + [\delta + \epsilon \cos(\nu t) + \epsilon' \cos^2(\nu t)] \langle \hat{x} \rangle = 0,
    \label{eq:13}
\end{align}
where $\delta = p (p + 2 c'_2)$, $\epsilon = 2 \epsilon_0 p (p + c'_2)$, and $\epsilon' = \epsilon_0^2 p^2$. Similarly, the dynamics under the Hamiltonian Eq.~(\ref{eq:4}) with a periodically driving quadratic Zeeman energy $q [1 + \epsilon_0 \cos(\nu t)]$, which is the same as the manipulation of Chapman's group, can be also mapped to Eq.~(\ref{eq:13}), but with coefficients $\delta = q (q + 4 c'_2)$, $\epsilon = 2 \epsilon_0 q (q + 2 c'_2)$, and $\epsilon' = \epsilon_0^2 q^2$~\cite{Luigi2013Spin}.

We notice that Eq.~(\ref{eq:13}) is exactly same as Eq.~(\ref{eq:1}) if one assumes $c'_2 = 0$. It implies that generalized parametric resonance happens under the Hamiltonian Eq.~(\ref{eq:6}) even for $c'_2 = 0$. However, this does not actually happen because the system just evolves as a Larmor precession under the Hamiltonian $p(t) \hat{J}_x$ and the oscillation amplitude of $\langle \hat{J}_z \rangle$ is conserved. This is due to the intrinsic difference between Eq.~(\ref{eq:1}) and the dynamics under Hamiltonian $p(t) \hat{J}_x$. The phase space of a classical harmonic oscillator lies in a two dimensional infinite plane, while that of a Larmor precession just follows a one dimensional circle. In order to extend one dimension to two dimensions, we have to break the $SU(1)$ symmetry by applying a term which does not commute with $\hat{J}_x$. Without loss of generality, the nonlinear term $\hat{J}_z^2$ satisfies the condition and keeps the oscillator harmonic. Of course, other terms  $\hat{J}_z^n$ with $n \neq 2$ also satisfy the condition but the oscillator becomes anharmonic.

In the regime of $p \gg |c'_2|$ but the nonlinear term still plays a role, Eq.~(\ref{eq:13}) is approximately same as Eq.~(\ref{eq:1}). In fact, the generalized parametric resonance in experiments would appear if $p$ is a dozen times as large as $|c'_2|$.

\section{Generalized Parametric resonances}
\label{sec:4}

To demonstrate the mapping process is correct and the dynamics under the Hamiltonians in Eq.~(\ref{eq:4}), Eq.~(\ref{eq:6}) and Eq.~(\ref{eq:7}) are consistent with the stability chart of GME, we numerically calculate the dynamics of $\langle \hat{J}_x \rangle$ under Hamiltonian Eq.~(\ref{eq:6}), and $\langle \hat{a}_0^\dagger \hat{a}_0 \rangle$ under Hamiltonian Eq.~(\ref{eq:4}). For simplicity, we show the results of $\langle \hat{J}_x \rangle$ for a ferromagnetic system in the main text. The results of $\langle \hat{J}_x \rangle$ for an antiferromagnetic system and the results of $\langle \hat{a}_0^\dagger \hat{a}_0 \rangle$ for a ferromagnetic system are shown in the Appendix.~\ref{sec:appc}. Typically, for a ferromagnetic $^{87}$Rb spinor BEC, $c'_2$ and $p$ are set to be $- 7$ Hz and $70$ Hz, respectively. We observe the dynamics of $\langle \hat{J}_x \rangle$ under the Hamiltonian Eq.~(\ref{eq:6}) with different driving strengths and frequencies. We consider two regimes where the collective spin dynamics is essentially captured by the GME Eq.~(\ref{eq:1}).

\begin{figure}[t]
  \centering
  \includegraphics[width = 3.3in]{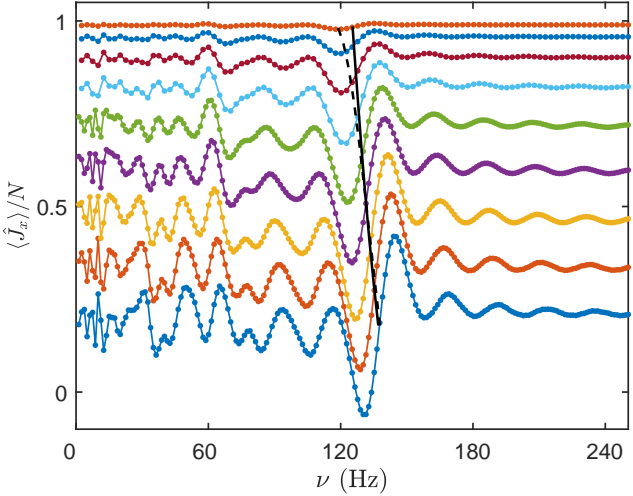}
  \caption{The polarizations $\langle \hat{J}_x \rangle$ at $45.5$ ms for different initial polarizations and driving frequencies. The initial polarization $\langle \hat{J}_x \rangle$ increases from bottom colored dotted line to top. The parameters are $c'_2 = - 7$ Hz, $p = 70$ Hz, and $\epsilon_0 = 0.4$. The total particle number is $2000$. The black solid (dashed) line shows the theoretical prediction for $2 \omega_1$ ($2 \omega_2$) resonance with $\omega_1$ ($\omega_2$) the effective oscillation frequency.}
  \label{fig:2}
\end{figure}

First, we consider a small driving strength $\epsilon_0 = 0.4$ to check whether there exists a parametric resonance. Numerical results for $\langle \hat{J}_x \rangle$ at time $45.5$ ms for different initial polarizations and different driving frequencies are shown in Fig.~\ref{fig:2}. As shown in Fig.~\ref{fig:2}, we find clear parametric resonances near $\nu = \omega_0$ and $\nu = 2 \omega_0$, corresponding to $\delta / \nu^2 = 1$ and $\delta / \nu^2 = 0.25$, respectively. Away from these resonances, the polarizations $\langle \hat{J}_x \rangle$ almost stay at their original values. The numerical results in Fig.~\ref{fig:2} are consistent with the stability chart with small $\epsilon$ in Fig.~\ref{fig:1}.

After careful check, however, we find the resonant frequency does not equal exactly $2 \omega_0 / n$ with integer $n$, except for $\langle \hat{J}_x \rangle / N \approx 1$. The reason is that the mapping process in Sec.~\ref{sec:3} requires $\varepsilon \ll x \ll 1$, which limits the frequency $\omega_0$ in Eq.~(\ref{eq:11}) to be in the vicinity of $\langle \hat{J}_x \rangle / N \approx 1$. As the value of $\langle \hat{J}_x \rangle / N$ deviates from 1, the oscillator becomes anharmonic because of the higher order terms. To incorporate the effect of anharmonicity, one usually employs an alternative harmonic oscillator but with a revised effective oscillation frequency. Such an effective oscillation frequency can be obtained by solving the equations of motion in Heisenberg picture or solving an elliptic equation in the mean field approximation.

The equations of motion in the Heisenberg picture for the angular momenta along $y$ and $z$ axes are,
\begin{align}
    \dot{\hat{J}}_z &= - p \hat{J}_y, \nonumber \\
    \dot{\hat{J}}_y &= p \hat{J}_z + \frac{c'_2}{N} (\hat{J}_x \hat{J}_z + \hat{J}_z \hat{J}_x).
\end{align}
When $p$ is much larger than $c'_2$, we regard $\langle \hat{J}_x \rangle$ as a constant because of its small oscillation amplitude. By defining $\gamma \equiv \langle \hat{J}_x \rangle / N$, the above equations are analytically solvable,
\begin{align}
    \hat{J}_z(t) &= \cos(\omega_1 t) \hat{J}_z(0) - \frac{p}{\omega_1} \sin(\omega_1 t) \hat{J}_y(0), \nonumber \\
    \hat{J}_y(t) &= \frac{\omega_1}{p}\sin(\omega_1 t) \hat{J}_z(0) + \cos(\omega_1 t) \hat{J}_y(0),
\end{align}
where the effective frequency is
\begin{align}
    \omega_1 = \sqrt{p (p + 2 \gamma c'_2)}.
    \label{eq:16}
\end{align}

In the mean field approximation, the Hamiltonian in Eq.~(\ref{eq:6}) becomes~\cite{Zhang2005Coherent, Zhang2015Coherent},
\begin{align}
    H / N = c'_2 f_z^2 - p f_x,
\end{align}
where $f_x = \langle \hat{J}_x \rangle / N$, $f_z = \langle \hat{J}_z \rangle / N$. By treating the condensate spin as a classical spin, which rotates in an effective magnetic field $(B_x, B_y, B_z) = (- p, 0, 2 c'_2 f_z)$, we obtain the following equation of motion, $\dot{f}_x = d f_x / d t = - 2 c'_2 f_z f_y$. By utilizing further the relations $f_y^2 = f^2 - f_x^2 - f_z^2$ and $f_z^2 = (E_{xz} + \Omega f_x) / c'_2$ with $E_{xz}$ the conserved energy, we find a closed equation of motion for $f_x$,
\begin{align}
    \dot{f}_x^2 = 4 c'_2 (E_{xz} + p f_x) (f^2 - f_x^2) - 4 (E_{xz} + p f_x)^2,
\end{align}
where $f$ and $E_{xz}$ are determined by the initial condition. It is straightforward to compute the oscillation period for $f_x$,
\begin{align}
    T = \oint \frac{1}{\dot{f}_x} d f_x = \frac{2}{\sqrt{- p c'_2}} \frac{K \left( \frac{x_2 - x_1}{x_3 - x_1} \right)}{\sqrt{x_3 - x_1}},
\end{align}
where $K(k)$ is the complete elliptic integral of the first kind, and $x_{j = 1, 2, 3}$ are the roots of $\dot{f}_x = 0$, with $x_1 \leqslant x_2 \leqslant x_3$. The numerical results of $x_{j = 1, 2, 3}$ can be found in Appendix.~\ref{sec:appb}. We find that $(x_2 - x_1) / (x_3 - x_1) \approx 0$, so $K(k \approx 0) \approx \pi / 2$ and $\sqrt{- p c'_2} \sqrt{x_3 - x_1} \approx \sqrt{p (p + 2 c'_2 \theta)}$, with $\theta \ll 1$ the polar angle of the total mean spin. Consequently, we obtain the effective frequency of the oscillator,
\begin{align}
    \omega_2 = \sqrt{p (p + 2 \theta c'_2)}.
    \label{eq:20}
\end{align}

Both Eq.~(\ref{eq:16}) and Eq.~(\ref{eq:20}) show that the effective oscillation frequency increases as $\langle \hat{J}_x \rangle$ decreases. According to the above approximations, Eq.~(\ref{eq:16}) is valid in the regime of $\langle \hat{J}_x \rangle / N \approx 1$ while Eq.~(\ref{eq:20}) is valid for $\langle \hat{J}_x \rangle / N \approx 0$. However, after replacing $\gamma$ with its average in one period $\gamma \equiv  \omega \int_0^{1 / \omega} dt \langle \hat{J}_x \rangle / N$, we find Eq.~(\ref{eq:16}) is in fact valid for the whole regime. As shown in Fig.~\ref{fig:2}, the theoretical predictions for resonant frequencies according to Eq.~(\ref{eq:16}) agree well with the numerical results.

\begin{figure}[t]
    \centering
    \includegraphics[width = 3.2in]{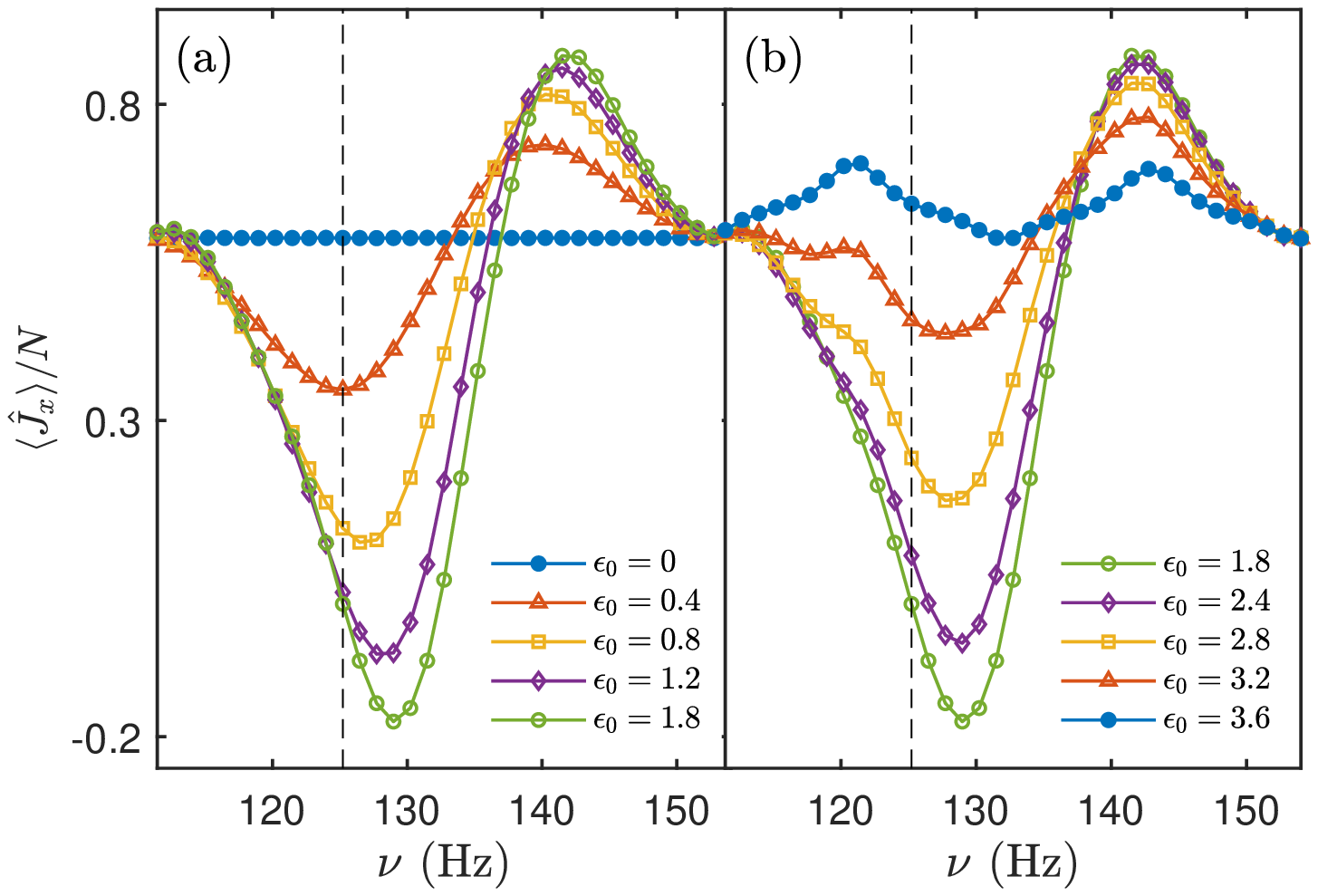}\vspace{0.5cm}
    \includegraphics[width = 3.2in]{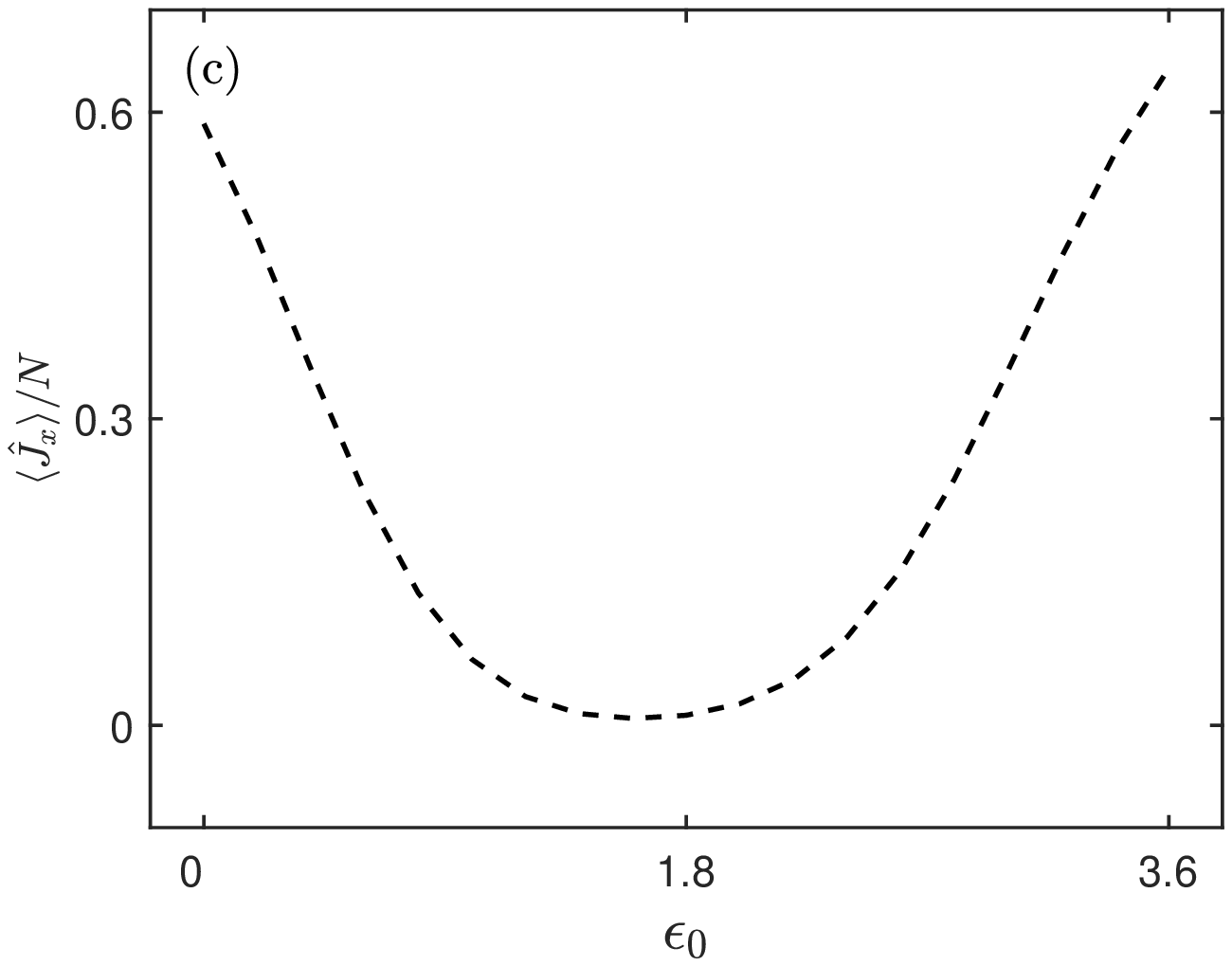}
    \caption{(a, b) Dependence of the polarizations $\langle \hat{J}_x \rangle$ at $45.5$ ms on the driving frequency for different driving strengths. The initial polarization $\langle \hat{J}_x \rangle / N = 0.6$. The intersections of the black dashed line and colored lines with different symbols in (a, b) are shown in (c).}
    \label{fig:3}
\end{figure}

Second, we consider the strong driving cases which are beyond the ME but still described well by the GME in the spin space under the Hamiltonian Eq.~(\ref{eq:6}). We choose $\epsilon_0 \in [0, 4]$, including the large driving strengths, and $\nu = 2 \omega_0$, corresponding to $\epsilon / \nu^2 \in [0, 2]$ and $\delta / \nu^2 = 0.25$, respectively. The numerical results for polarizations $\langle \hat{J}_x \rangle$ at $45.5$ ms are shown in Fig.~\ref{fig:3}. As shown in Fig.~\ref{fig:3}(a) and (b), we find the amplitude of the parametric excitation does not always increase as the driving strength increases in the regime $\epsilon_0 \in [0, 4]$. The intersections of the black dashed line $\nu = 2 \omega_0$ and the colored lines, which manifest the excitation amplitude, are shown in Fig.~\ref{fig:3}(c). Obviously, the excitation amplitude reaches its extremum near $\epsilon_0 \approx 1.6$, corresponding to $\epsilon / \nu^2 \approx 0.8$, a value close to the transition point in Fig.~\ref{fig:1}. The decrease of the resonance amplitude as $\epsilon_0$ further increases ($\epsilon_0 > 1.6$) implies the agreement between the numerical dynamics under Hamiltonian Eq.~(\ref{eq:6}) and the prediction of the GME, instead of the ME which predicts unstable dynamics. This result is interesting because the system does not absorb energy when the external driving strength increases to a value higher than a critical point. Similar results of $\langle \hat{J}_x \rangle$ for the antiferromagnetic $^{23}$Na system in spin space and $\langle \hat{a}_0^\dagger \hat{a}_0 \rangle$ for the ferromagnetic $^{87}$Rb system in nematic space can be found in Appendix.~\ref{sec:appc}. Both of the numerical calculations show a similar behavior as depicted in Fig.~\ref{fig:3}. Furthermore, the numerical calculations of the ferromagnetic $^{87}$Rb system in nematic space not only well explain the data of Chapman's group with a small driving strength, but predict the dynamics in the regime of large driving strength.

The appearance of a saddle point in Fig.~\ref{fig:3}(c) can be well understood as follows. After a unitary transformation $\hat{U} = \exp\{- i \int_0^t p (1 + \epsilon_0 \cos(\nu t)) \hat{J}_x d\tau\}$, the Hamiltonian in Eq.~(\ref{eq:6}) becomes,
\begin{align}
    \hat{H} = \frac{c'_2}{N} (\hat{J}_z \cos(\phi) - \hat{J}_y \sin(\phi))^2,
\end{align}
where $\phi = p (t + \epsilon_0 \sin(\nu t) / \nu)$. Then the Heisenberg equation for $\hat{J}_x$ is,
\begin{align}
    \dot{\hat{J}}_x = - \{ \hat{J}_z, \hat{J}_y \} \cos(2 \phi) - \hat{J}_z^2 \sin(2 \phi) + \hat{J}_y^2 \sin(2 \phi).
\end{align}
According to the Jacobi-Anger expansions,
\begin{align}
    \cos(z \sin(\alpha)) &= \mathcal{J}_0 + 2 \sum_{n = 1}^{\infty} \mathcal{J}_{2 n}(z) \cos(2 n \alpha), \nonumber \\
    \sin(z \sin(\alpha)) &= 2 \sum_{n = 1}^\infty \mathcal{J}_{2 n - 1}(z) \sin((2 n - 1) \alpha),
\end{align}
by taking further the rotating wave approximation, we obtain
\begin{align}
    \langle \psi | \hat{J}_x(t) | \psi \rangle =& \frac{c'_2}{N} \mathcal{J}_1\left(\frac{2 p \epsilon_0}{\nu}\right) \langle \psi | \{ \hat{J}_z(0), \hat{J}_y(0) \} \nonumber \\
    &- \hat{J}_z^2(0) + \hat{J}_y^2(0) | \psi \rangle t+ \langle \psi | \hat{J}_x(0) | \psi \rangle.
\end{align}
The above equation shows that the maximum of excitation amplitude happens at $2 p \epsilon_0 / \nu = 1.84$, based on the property of Bessel function. One immediately finds $\epsilon_0 = 1.65$ when $\nu = 2 \omega_0$. 

Finally, we would like to stress that the amplitude of $x$ in the GME may increase to infinity as time evolves; however, in our proposal for a spin-1 BEC, although the dynamics of the polarization can be approximately mapped to the GME, the amplitude of $\langle \hat{J}_x \rangle$ is physically finite, which not only results from the finite time evolution but is due to the intrinsically finite phase space of a spin or nematic spin.

\section{Conclusion}
\label{sec:5}

In conclusion, we have proposed a GME and mapped two different dynamics (one in nematic space and one in spin space) for the spin-1 system to this GME. The results of numerical calculations for dynamics under the Hamiltonian agree well with the stability chart of the GME. The unstable dynamics are called generalized parametric resonance. Based on our proposal, we not only explain the experimental results of Chapman's group in nematic space with a small driving strength, but also predict the behaviors in the regime of large driving strength. The dynamics in spin space we propose can be fine tuned and is easily to be implemented in current experimental conditions.

\section{Acknowledgments}

We thanks Qi Liu for discussions on the experimental details. The project is supported by the China Postdoctoral Science Foundation Grant No. 2020M680497, the National Natural Science Foundation of China Grant Nos. 91836101 and U1930201.

\newpage

\appendix

\renewcommand{\theequation}{\Alph{section}.\arabic{equation}}
\renewcommand{\thefigure}{\Alph{section}\arabic{figure}}

\setcounter{equation}{0}
\setcounter{figure}{0}

\section{The stability chart of the GME}
\label{sec:appa}

There are two ways to obtain the stability chart of the GME shown in Eq.(1) in the main text~\cite{Kovacic2018Mathieu}. First, we transform the second order differential equation to two first order differential equations, which are generally represented in the following,
\begin{align}
    \dot{\bm{x}} = \bm{A}(t) \bm{x},
\end{align}
where $\bm{x}$ is a vector, and $\bm{A}$ is a matrix. For a given periodic driving, $\bm{A}(t) = \bm{A}(t + T)$, with $T$ the driving period, we can obtain the effective evolution matrix in one period as,
\begin{align}
    \bm{A}_{\text{eff}} = \frac{\log(\mathcal{T} e^{\int_0^T \bm{A}(t) dt})}{T},
\end{align}
with $\mathcal{T}$ the time ordering operator. Then, if the eigenvalues of $\bm{A}_{\text{eff}}$ are purely imaginary numbers, the dynamics under the GME is stable; otherwise, it is unstable because the position or the velocity will deviate far away from its initial condition. The numerical results are shown in Fig.~\ref{fig:1}, and the shaded orange (light cyan) areas represent the unstable regions for the GME (ME).

Second, according to the perturbation theory, we expand $\delta$ in a power series in $\epsilon$,
\begin{align}
    \delta = \frac{n^2}{4} + \delta_1 \epsilon + \delta_2 \epsilon^2,
    \label{eq:A3}
\end{align}
and expand the solution in the form of a Fourier series,
\begin{align}
    x(t) = \sum_{n = 0}^\infty \left( a_n \cos \frac{n t}{2} + b_n \sin \frac{n t}{2} \right).
    \label{eq:A4}
\end{align}
Then we substitute Eq.~(\ref{eq:A4}) into Eq.~(\ref{eq:1}), simplifying the trigonometric functions and collecting terms with the same frequency. Finally, we obtain
\begin{align}
    &\sum_{n = 0}^\infty \left( \delta - \frac{n^2}{4} + \frac{\epsilon^2}{8 \delta} \right) a_n \cos \left( \frac{n t}{2} \right) \nonumber \\
    &+ \sum_{n = 0}^\infty \frac{\epsilon}{2} a_n \left[ \cos \left( \frac{n + 2}{2} t \right) + \cos \left( \frac{n - 2}{2} t \right) \right] \nonumber \\
    &+ \sum_{n = 0}^\infty \frac{\epsilon^2}{8 \delta} a_n \left[ \cos \left( \frac{n + 4}{2} t \right) + \cos \left( \frac{n - 4}{2} t \right) \right] = 0, \nonumber \\
    &\sum_{n = 0}^\infty \left( \delta - \frac{n^2}{4} + \frac{\epsilon^2}{8 \delta} \right) b_n \sin \left( \frac{n t}{2} \right) \nonumber \\
    &+ \frac{\epsilon}{2} b_n \left[ \sin \left( \frac{n + 2}{2} t \right) + \sin \left( \frac{n - 2}{2} t \right) \right] \nonumber \\
    &+ \frac{\epsilon^2}{8 \delta} b_n \left[ \sin \left( \frac{n + 4}{2} t \right) + \sin \left( \frac{n - 4}{2} t \right) \right] = 0.
\end{align}
The above equations give four sets of algebraic equations on the coefficients $a_{\text{even}}, b_{\text{even}}, a_{\text{odd}}, b_{\text{odd}}$. For a nontrivial solution the determinants must vanish. This gives the following four infinite determinants,
\begin{widetext}
    \begin{align}
        a_{\text{even}} :
        \left|
        \begin{array}{ccccc}
            \delta + \epsilon^2 / 8 \delta & \epsilon & \epsilon^2 / 4 \delta & 0 & \\
            \epsilon / 2 & \delta - 1 + \epsilon^2 / 4 \delta & \epsilon / 2 & \epsilon^2 / 8 \delta & \cdots \\
            \epsilon^2 / 8 \delta & \epsilon / 2 & \delta - 4 + \epsilon^2 / 8 \delta & \epsilon / 2 & \\
              &  & \cdots &  &
        \end{array}
        \right|
        = 0, \nonumber \\
        b_{\text{even}} :
        \left|
        \begin{array}{ccccc}
            \delta - 1 & \epsilon / 2 & \epsilon^2 / 8 \delta & 0 & \\
            \epsilon / 2 & \delta - 4 + \epsilon^2 / 8 \delta & \epsilon / 2 & \epsilon^2 / 8 \delta & \cdots \\
            \epsilon^2 / 8 \delta & \epsilon / 2 & \delta - 9 + \epsilon^2 / 8 \delta & \epsilon / 2 & \\
              &  & \cdots &  &
        \end{array}
        \right|
        = 0, \nonumber \\
        a_{\text{odd}} :
        \left|
        \begin{array}{ccccc}
            \delta - 1 / 4 + \epsilon / 2 + \epsilon^2 / 8 \delta & \epsilon / 2 + \epsilon^2 / 8 \delta & \epsilon^2 / 8 \delta & 0 & \\
            \epsilon / 2 + \epsilon^2 / 8 \delta & \delta - 9 / 4 + \epsilon^2 / 8 \delta & \epsilon / 2 & \epsilon^2 / 8 \delta & \cdots \\
            \epsilon^2 / 8 \delta & \epsilon / 2 & \delta - 25 / 4 + \epsilon^2 / 8 \delta & \epsilon / 2 & \\
              &  & \cdots &  &
        \end{array}
        \right|
        = 0, \nonumber \\
        b_{\text{odd}} :
        \left|
        \begin{array}{ccccc}
            \delta - 1 / 4 - \epsilon / 2 + \epsilon^2 / 8 \delta & \epsilon / 2 - \epsilon^2 / 8 \delta & \epsilon^2 / 8 \delta & 0 & \\
            \epsilon / 2 - \epsilon^2 / 8 \delta & \delta - 9 / 4 + \epsilon^2 / 8 \delta & \epsilon / 2 & \epsilon^2 / 8 \delta & \cdots \\
            \epsilon^2 / 8 \delta & \epsilon / 2 & \delta - 25 / 4 + \epsilon^2 / 8 \delta & \epsilon / 2 & \\
              &  & \cdots &  &
        \end{array}
        \right|
        = 0.
        \label{eq:A6}
    \end{align}
\end{widetext}
Each of these four determinants represents a functional relationship between $\delta$ and $\epsilon$, which plots a set of transition curves in the $\delta$-$\epsilon$ plane. By setting $\epsilon = 0$, it is easy to obtain where the associated curves intersect the $\delta$-axis, i.e., $\delta = n^2 / 4$ with $n = 0, 1, 2, \cdots$.
Then we substitute Eq.~(\ref{eq:A3}) with $n = 2$ into $a_{\text{even}}, b_{\text{even}}$ in Eq.~(\ref{eq:A6}). Expanding a $3 \times 3$ truncation of Eq.~(\ref{eq:A6}), we get
\begin{align}
    \delta_1 = 0, \quad \delta_2 = \frac{1}{6} \quad \text{for} \quad a_{\text{even}}, \nonumber \\
    \delta_1 = 0, \quad \delta_2 = - \frac{1}{12} \quad \text{for} \quad b_{\text{even}}.
\end{align}
Similarly, substituting Eq.~(\ref{eq:A3}) with $n = 1$ into $a_{\text{odd}}, b_{\text{odd}}$ in Eq.~(\ref{eq:A6}), we obtain
\begin{align}
    \delta_1 = - \frac{1}{2}, \quad \delta_2 = - \frac{5}{8} \quad \text{for} \quad a_{\text{odd}}, \nonumber \\
    \delta_1 = \frac{1}{2}, \quad \delta_2 = - \frac{5}{8} \quad \text{for} \quad b_{\text{odd}}.
\end{align}
The orange dashed lines in Fig.~\ref{fig:1} show the results of this perturbation calculations for the GME. Meanwhile, the light cyan solid lines are the perturbation results for the ME.

According to the numerical and perturbation theory's results, when $\delta$ is far away from $0$ and $\epsilon$ is small, we find the unstable regions of the GME moves to left comparing with the unstable regions of the ME. This can be understood by approximately rewriting Eq.~(\ref{eq:1}) as $\ddot{x} + \left[ \tilde{\delta} + \epsilon \cos(t) \right] = 0$ with $\tilde{\delta} = \delta + \epsilon^2 / (8 \delta)$. So, as $\epsilon$ increases, a smaller $\delta$ plays the same role as a bigger $\tilde{\delta}$. Besides, we find there are some narrow bands near $\delta = 0$, which can be understood by approximately rewriting Eq.~(\ref{eq:1}) as $\ddot{x} + \left[ \epsilon^2 / (8 \delta) + \epsilon^2 / (8 \delta) \cos(2 t) \right] = 0$. Obviously, the resonance occurs at the condition $\delta = \epsilon^2 / (8 n^2)$. The appearance of narrow bands shows the difference between the GME we proposed in this paper and another model $\ddot{x} + \left[ \delta + \epsilon_1 \cos(t) + \epsilon_2 \cos(2t) \right] = 0$ with $\epsilon_2 \gg \epsilon_1$. More importantly, there exists a region in the vicinity of $\delta \approx 1 / 4, \epsilon \gtrsim  1$ that is stable for the GME but unstable for the ME. This region can be used to benchmark the dynamics under Eq.~(\ref{eq:4}) and Eq.~(\ref{eq:6}) that are well described by the GME, but not by the ME.

\setcounter{equation}{0}
\setcounter{figure}{0}

\section{The roots of $\dot{f}_x = 0$}
\label{sec:appb}

\begin{figure}[t]
    \centering
    \includegraphics[width = 3.3in]{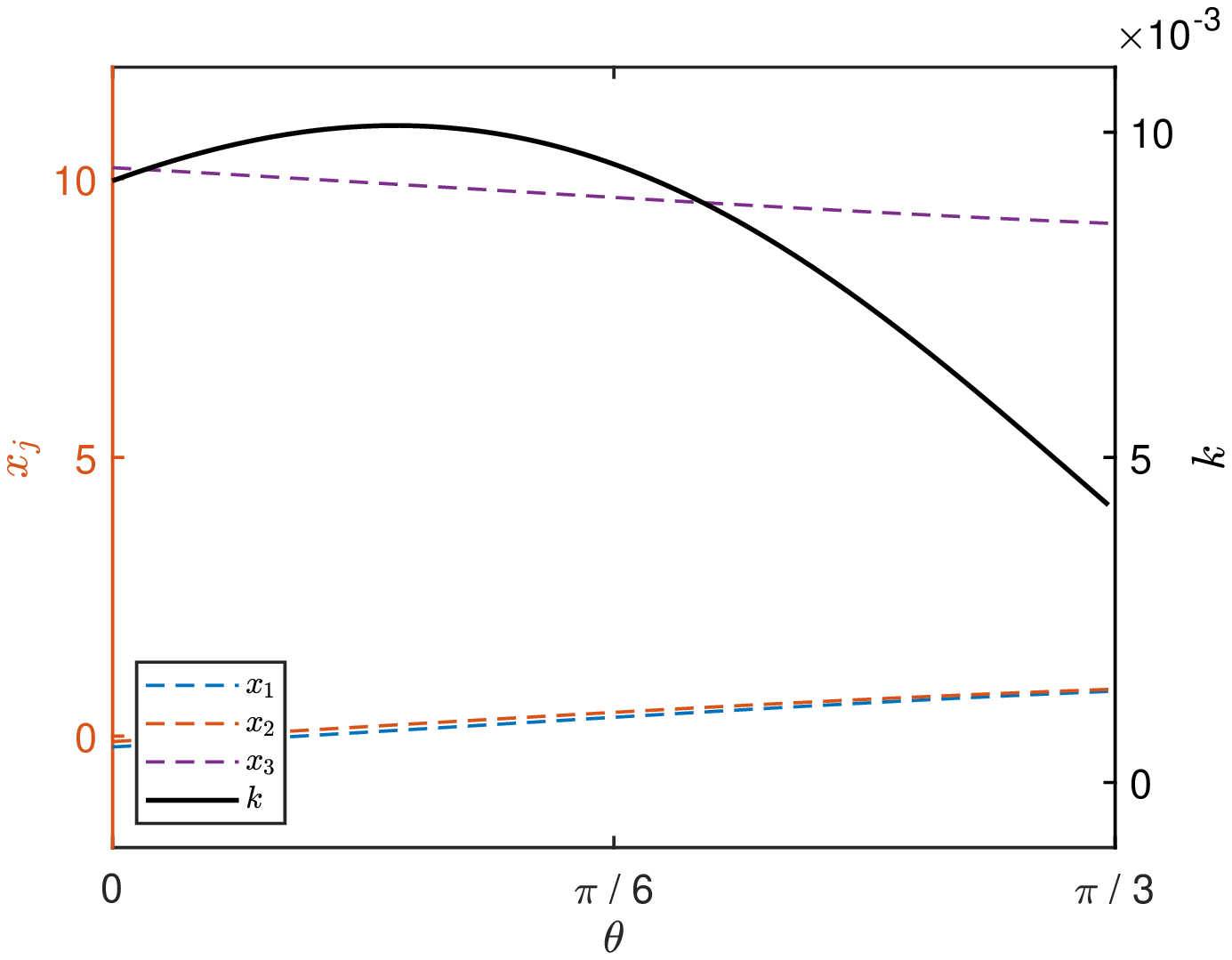}
    \caption{The colored dashed lines show the roots of $\dot{f}_x = 0$ varying with the polar angle $\theta$. The black solid line shows the variable of $k = (x_2 - x_1) / (x_3 - x_1)$ for $p / c'_2 = - 10$.}
    \label{fig:B1}
\end{figure}

The roots of $\dot{f}_x = 0$ are shown in the following,
\begin{align}
    x_1 &= \frac{- p + \sqrt{- 4 c'_2 E_{xz} + 4 {c'_2}^2 f^2 + p^2}}{2 c'_2}, \nonumber \\
    x_2 &= - \frac{E_{x z}}{p}, \nonumber \\
    x_3 &= \frac{- p - \sqrt{- 4 c'_2 E_{xz} + 4 {c'_2}^2 f^2 + p^2}}{2 c'_2}.
\end{align}
The numerical results for $x_{j = 1, 2, 3}$ and $k$ in the condition $p / c'_2 = - 10$ are shown in Fig.~\ref{fig:B1}. We find $k$ almost equals 0 for $\theta \in [0, \pi / 3]$.

\setcounter{equation}{0}
\setcounter{figure}{0}

\section{Dynamics under parameters including strong driving strengths for antiferromagnetic $^{23}$Na in spin space and ferromagnetic $^{87}$Rb in nematic space}
\label{sec:appc}

\begin{figure}[t]
    \centering
    \includegraphics[width = 3.3in]{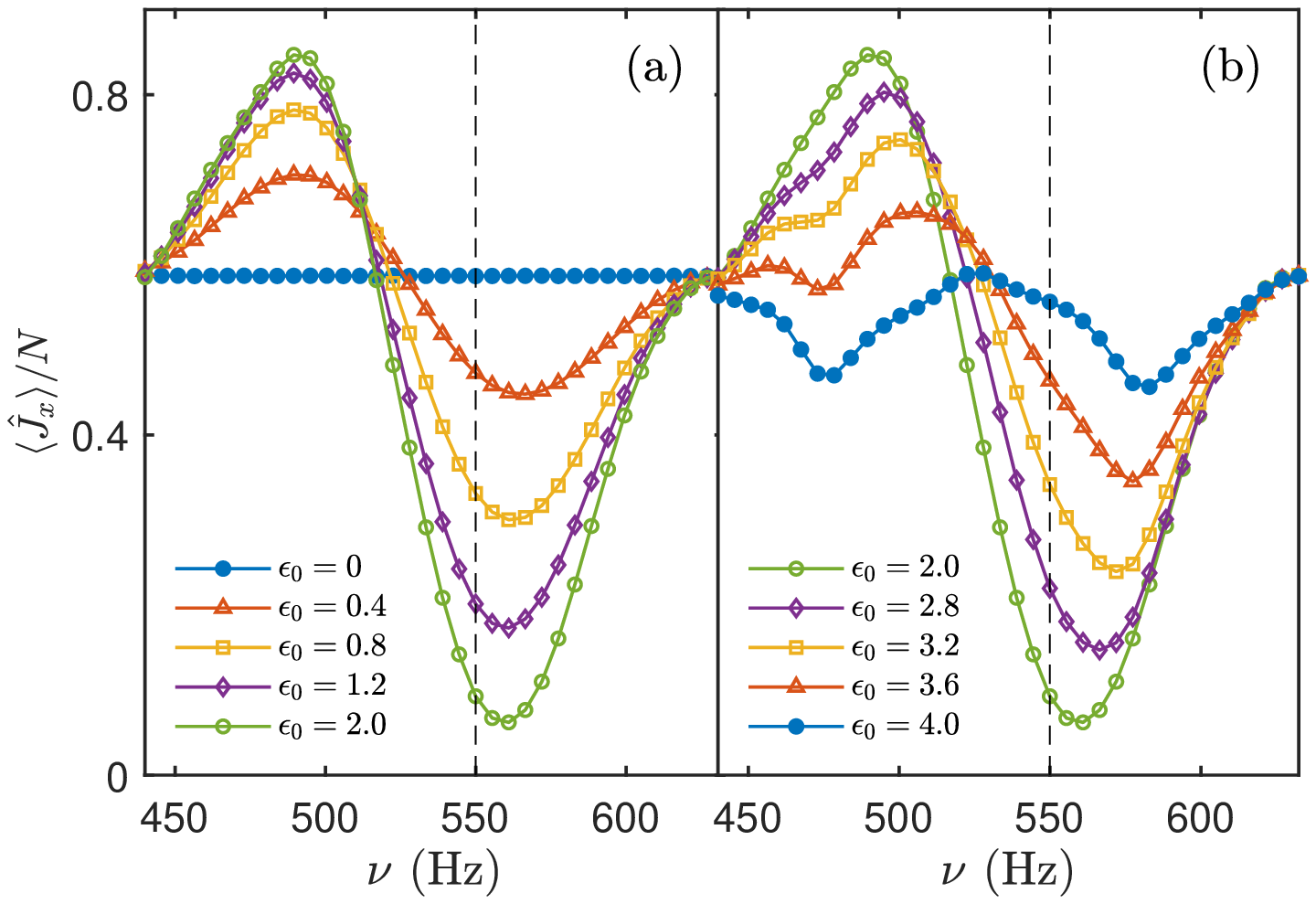}\vspace{0.5cm}
    \includegraphics[width = 3.3in]{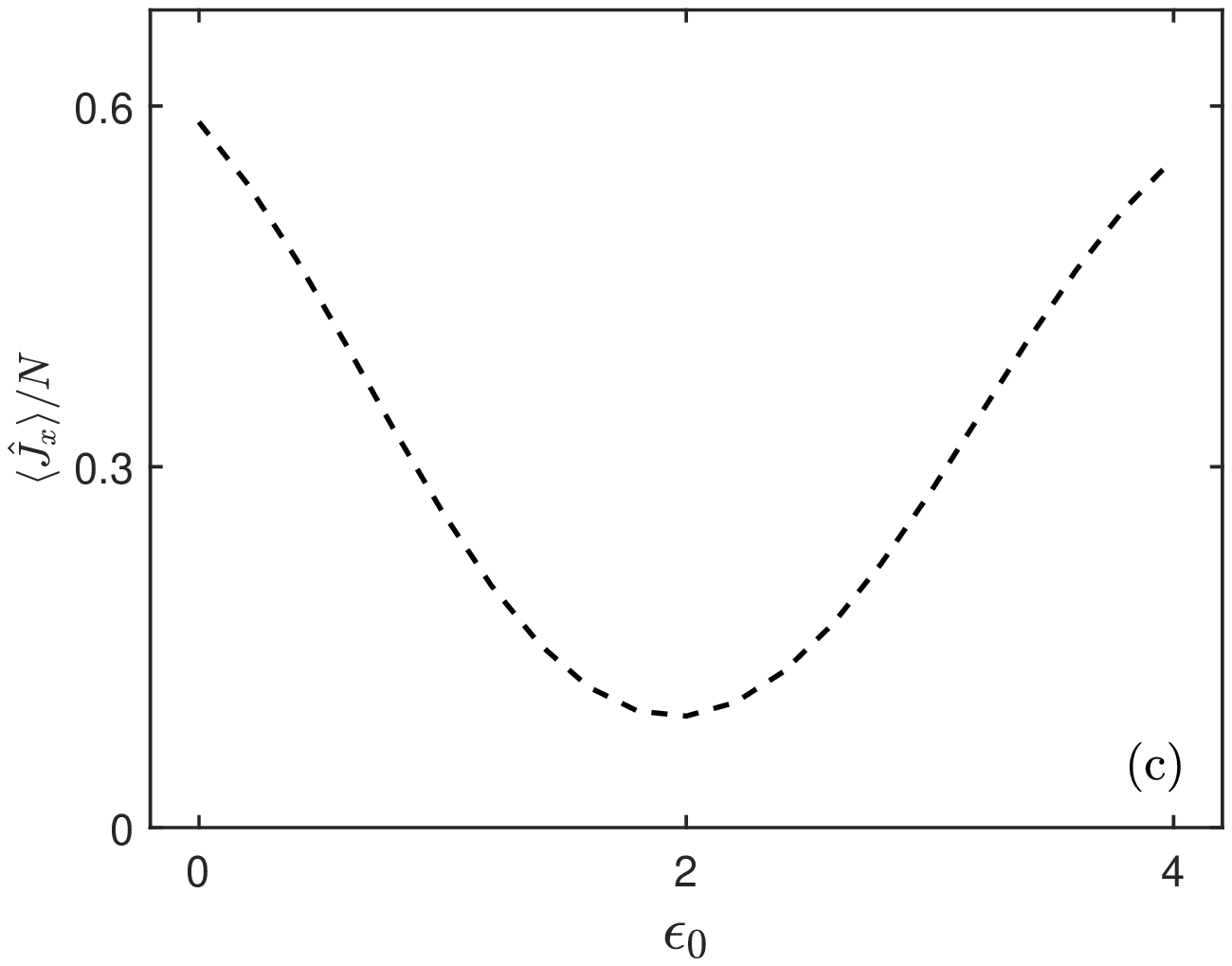}
    \caption{(a, b) Dependence of the polarizations $\langle \hat{J}_x \rangle$ at $12.7$ ms on the driving frequency for different driving strengths. The initial polarization $\langle \hat{J}_x \rangle / N = 0.6$. The parameters are $c'_2 = 25$ Hz, $p = 250$ Hz, $\epsilon_0 \in [0, 4]$. The intersections of the black dashed line ($\nu = 2 \omega_0$) and colored lines with different symbols in (a, b) are shown in (c).}
    \label{fig:C1}
\end{figure}

\begin{figure}[t]
    \centering
    \includegraphics[width = 3.3in]{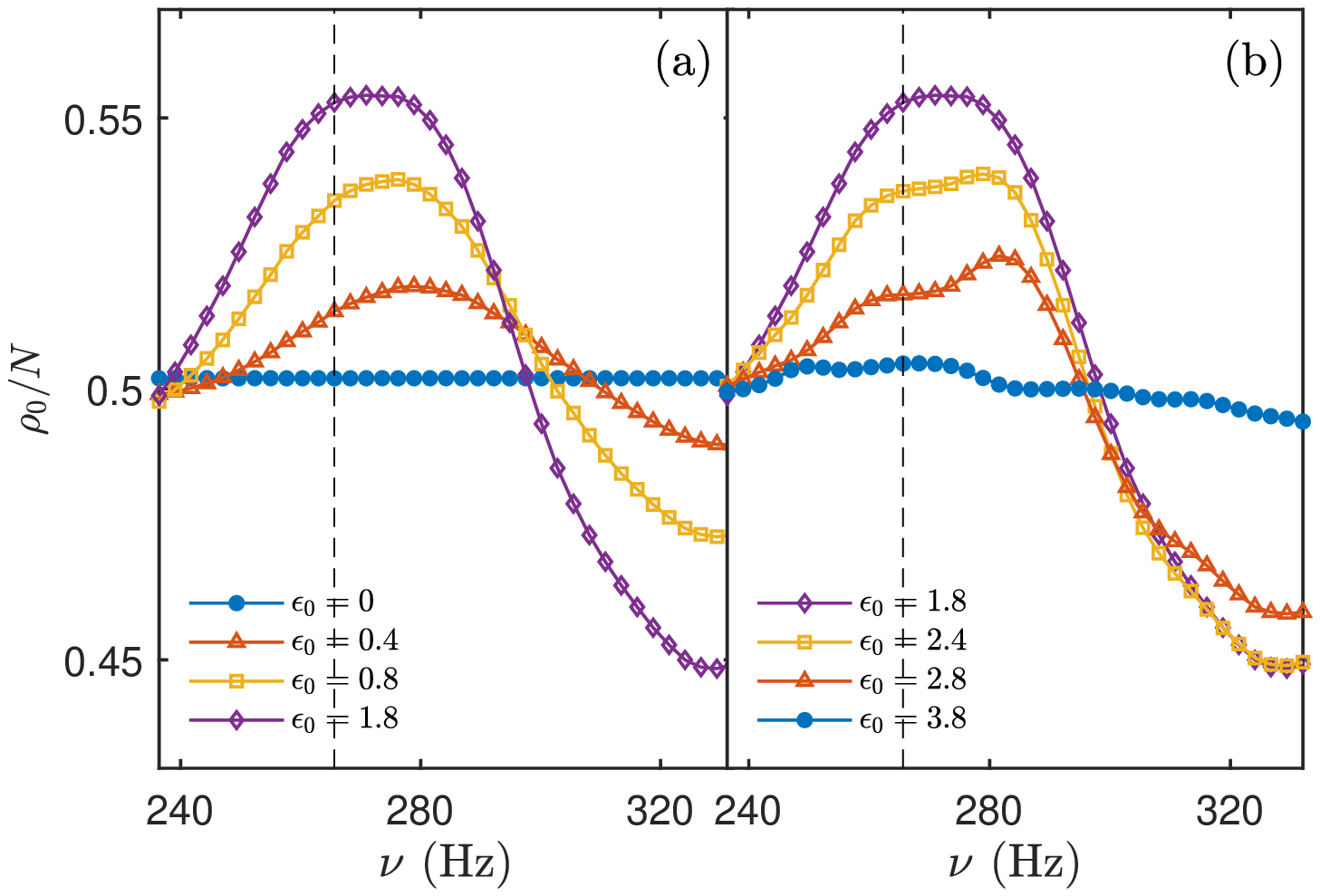}\vspace{0.5cm}
    \includegraphics[width = 3.3in]{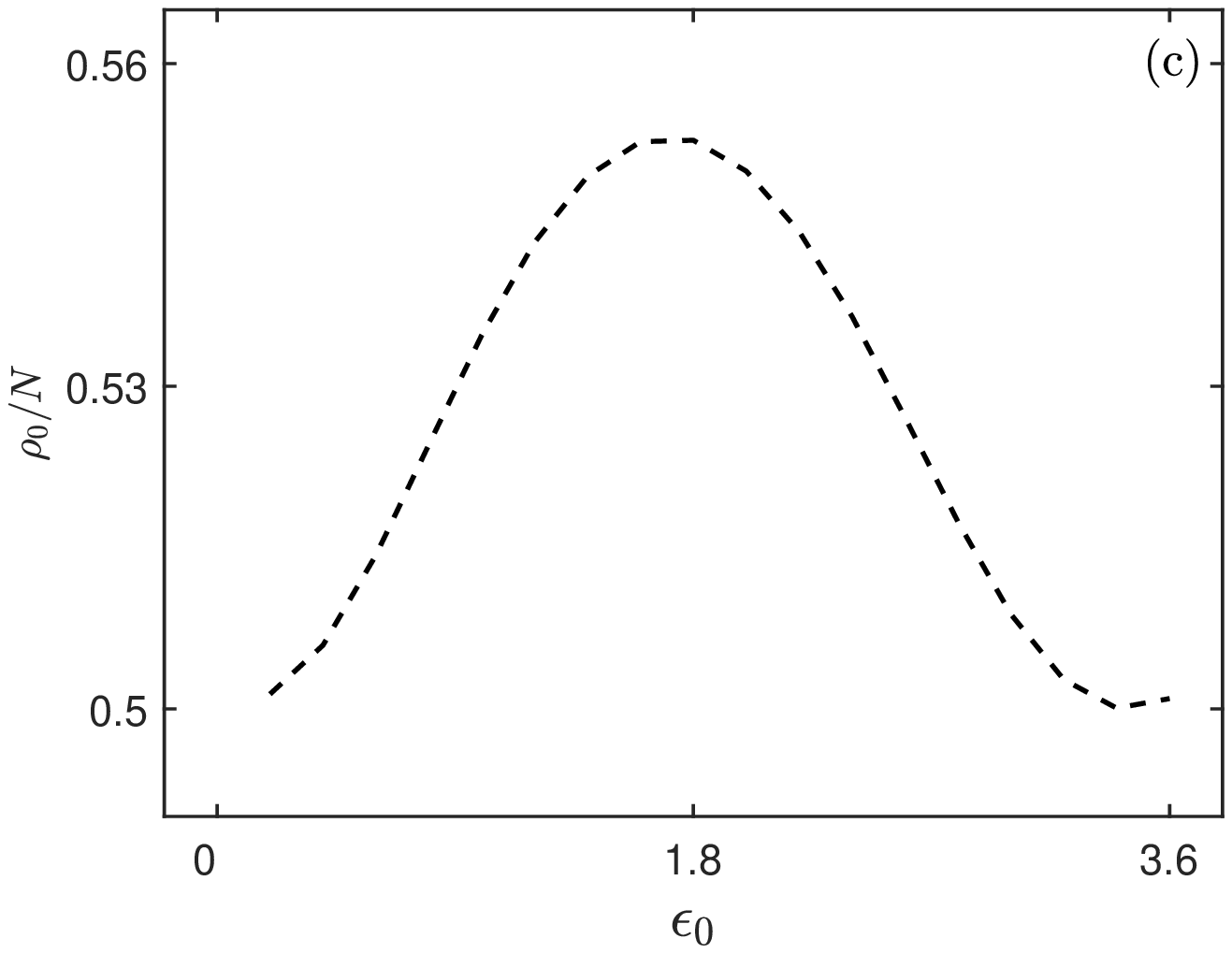}
    \caption{(a, b) Populations $\rho_0$ at $17.1$ ms for different driving strengths and different driving frequencies. The initial state is $\rho_0 / N = 0.5$, with total particle number $N = 200$. The typical values $c'_2 = - 7$ Hz, $q = 140$ Hz, $\epsilon_0 \in [0, 4]$. The intersections between black dashed line ($\nu = 2 \omega_0$) and colored lines with different symbols in (a, b) are shown in (c).}
    \label{fig:C2}
\end{figure}

The numerical calculations of $\langle \hat{J}_x \rangle$ for an antiferromagnetic $^{23}$Na BEC in spin space and $\langle \hat{\rho}_0 \rangle = \langle \hat{a}_0^\dagger \hat{a}_0 \rangle$ for a ferromagnetic $^{87}$Rb BEC in nematic space are shown in  Figs.~\ref{fig:C1} and~\ref{fig:C2}, respectively. These results agree well with the stability chart of the GME. Furthermore, as shown in Fig.~\ref{fig:C2}, the numerical results are consistent with the experimental data of Chapman's group in the small driving strength. However, the excitation amplitude of population indeed decreases as $\epsilon_0$ increases till to the regime $\epsilon_0 \gtrsim 1.7$. 

\setcounter{equation}{0}
\setcounter{figure}{0}

\section{Typical evolutions corresponding to Fig.~\ref{fig:3}}
\label{sec:appd}

\begin{figure}[t]
  \centering
  \includegraphics[width = 3.3in]{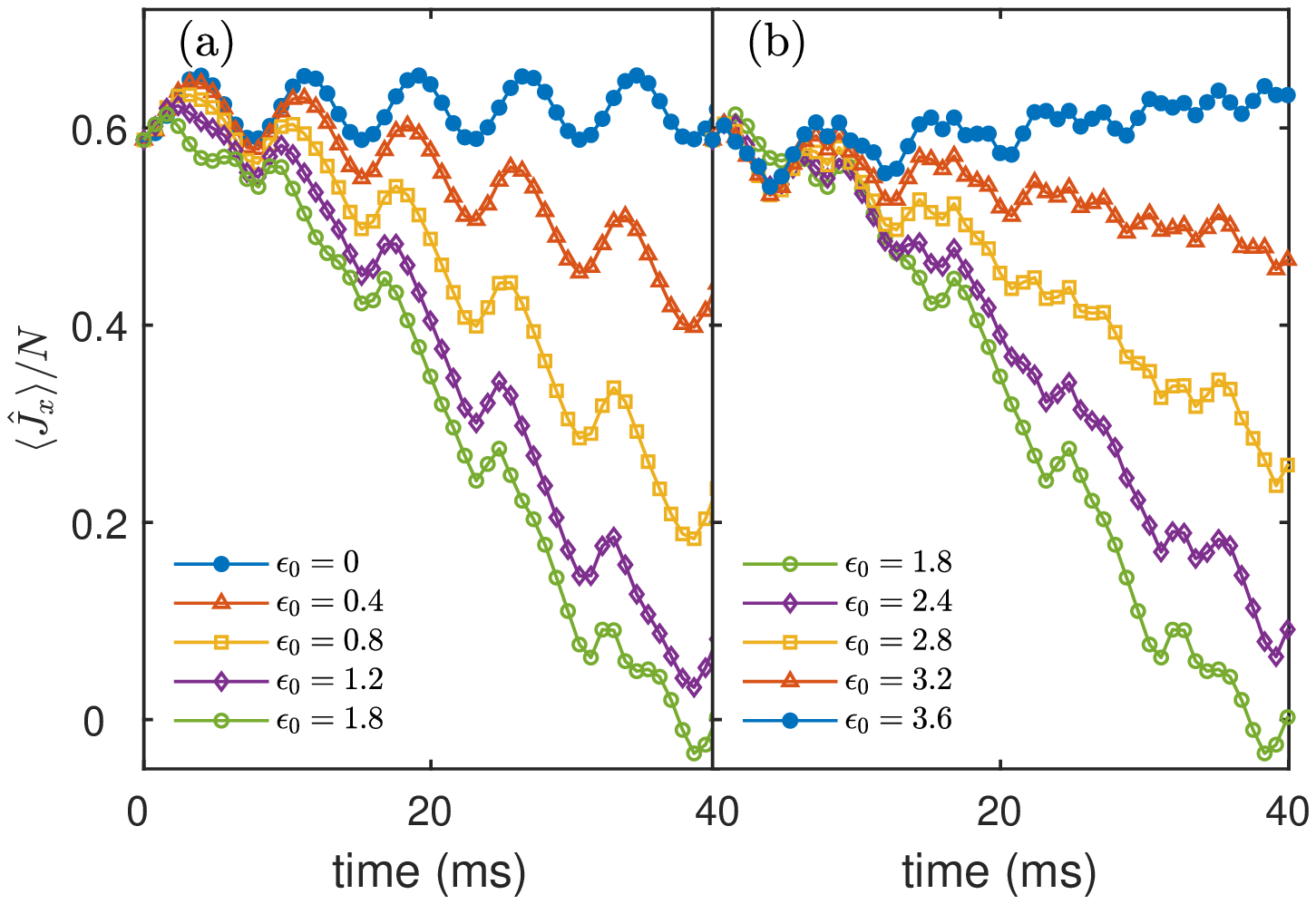}
  \caption{Typical time evolutions of $\langle \hat{J}_x \rangle$ under parameters $\epsilon_0 = 0, 0.4, \dots, 3.6$ and $\nu = 2 \omega_0$ for $^{87}$Rb ferromagnetic system. The other parameters are the same as parameters in Fig.~\ref{fig:3}.}
  \label{fig:D1}
\end{figure}

As a complementary material of Fig.~\ref{fig:3}, we show several typical evolutions under parameters $\epsilon_0 = 0, 0.4, \dots, 3.6$ and $\nu = 2 \omega_0$. The numerical results are shown in Fig.~\ref{fig:D1}. It shows that the oscillations, which are similar to the behavior in a simple harmonic oscillator, disappear in the regime of large driving strength.

% \bibliography{Generalized_Parametric_Resonance}

\end{document}